\documentclass[manuscript]{aastex62}
\usepackage{txfonts}
\usepackage{latexsym,bm}
\usepackage{url}
\usepackage{color}
\usepackage{colortbl}
\usepackage{multirow}
\usepackage{ulem} 

\definecolor{mygray}{gray}{.9}

\shorttitle{proton acceleration at the shock}

\shortauthors{Kong et al.}

\begin{document}

\title{Study of time evolution of the bend-over energy in the energetic
particle spectrum at a parallel shock}

\correspondingauthor{G. Qin}
\email{qingang@hit.edu.cn}

\author{F.-J. Kong}
\author[0000-0002-3437-3716]{G. Qin}
\author[0000-0002-5776-455X]{S.-S. Wu}

\affiliation{School of Science, Harbin Institute of Technology, Shenzhen,
518055, China; qingang@hit.edu.cn}

\author{L.-H. Zhang}
\author{H.-N. Wang}
\affiliation{Key Laboratory of Solar Activity, National Astronomical Observatories,
 Chinese Academy of Sciences, Beijing, 100012, China}

\author{T. Chen}
\affiliation{State Key Laboratory of Space Weather, National Space Science Center, 
Chinese Academy of Sciences, Beijing, 100190, China}

\author{P. Sun}
\affiliation{Department of Planetary Sciences and Astronomy, University of Arizona, 
Tucson, AZ 85721, USA}

\begin{abstract}
Shock acceleration is considered one of the most important mechanisms for 
the acceleration of astrophysical energetic particles. In this work, we 
calculate the trajectories of a large number of test charged particles 
accurately in a parallel shock with magnetic turbulence. We investigate the 
time evolution of the accelerated-particle energy spectrum in the downstream 
of the shock in order to understand the acceleration mechanism of energetic particles. 
\textbf{From simulation results we obtain power-law energy spectra with a bend-over energy, 
$E_0$, increasing with time.} With the particle mean acceleration time and mean momentum 
change during each cycle of the shock crossing from diffusive shock acceleration model (following Drury), 
a time-dependent differential equation for the maximum energy, $E_{acc}$, of particles 
accelerated at the shock, can be approximately obtained. We assume the theoretical 
bend-over energy as $E_{acc}$. It is found that the bend-over energy from simulations 
agrees well with the theoretical bend-over energy using the non-linear diffusion theory, 
NLGCE-F, in contrast to that using the classic quasi-linear theory (QLT).

\end{abstract}


\section{Introduction}
The collisionless shock acceleration of energetic particles, which is considered
to be one of the key problems to study the sources of solar energetic particles
(SEPs) and galactic cosmic rays (GCRs), has been studied by many scholars in
decades \citep[e.g.,][]{Fermi1949, Bell1978, Jokipii1982, Drury1983, Decker1988,
Krulls1994, Lee1996, ZankEA00, Bell2004, Sun2007, Florinski2008, LiGang2012,
WangEA12, QinEA13, QinEA2018, Zuo2013, QiEA17, KongEA2017}. 
Different physical mechanisms have been developed to explain the shock acceleration
processes. One mechanism is shock drift acceleration (SDA) 
\cite[]{Jokipii1982, Forman1985, Lee1996, Shapiro2003, GuoXY2014},\textbf{ which mainly
occurs at perpendicular or quasi-perpendicular shocks. The discontinuous tangential
component of the magnetic field across the shock results in the
gradient drift of the particles along the shock front in the direction 
as the electric field in the shock frame.} In this case, the energy gain for each gyro-cycle, $\Delta E$, can be shown
in the form of 
\begin{equation}
\Delta E= q \bm{E}\cdot\Delta \bm{x},
\end{equation}
where $\Delta\bm{x}$ is the particle displacement variance, and 
the drift electrostatic field in the shock frame, $\bm{E}$, is given by 
\begin{equation}
\bm{E} = -\bm{U}\times\bm{B},
\end{equation}
where $\bm{U}$ is the background bulk velocity. 
\textbf{Note that if particles have initial speeds that are comparable to the background flow speed,
they would move along the magnetic field lines and convect into the downstream. The SDA therefore
generally has an injection problem for low-energy particles \citep[e.g.,][]{Ellison1995,Jokipii1996}. 
However, low-energy particles can cross the shock repeatedly under large-scale magnetic fluctuations,
and are effectively injected into diffusive shock acceleration} \citep{Giacalone2005a,Giacalone2005b}.

\textbf{Another mechanism is first-order Fermi acceleration, which, together 
with SDA, is incorporated into the diffusive shock acceleration (DSA)
 \citep{Fermi1949, Bell1978, Drury1983, Forman1985, 
Kirk1987,Malkov2001, Amato2005, Amato2006}.  
In the first-order Fermi acceleration mechanism, the acceleration of particles is
derived from the relative motion 
of the scattering centers in the upstream and downstream of the shock.}
\citet{Drury1983} showed that for each cycle of the particle's crossing of the
shock front, the mean acceleration time $\Delta t$ is 
\begin{equation}
   \Delta t=\frac{4}{v}\left(\frac{\kappa_1}{U_1}+\frac{\kappa_2}{U_2}\right),
	\label{eqs:deltat}\\
\end{equation}
and the average momentum change is
\begin{eqnarray}
   \langle\Delta p\rangle&=&2p\int_0^1\frac{\mu(U_1-U_2)}{v}2\mu d\mu\nonumber\\
	{}&=&\frac{4}{3}\frac{U_1-U_2}{v}p,\label{eqs:deltap}
\end{eqnarray}
where $\kappa$ and $U$ represent the diffusion coefficient and background bulk velocity, 
respectively, the subscripts $1$ and $2$ indicate variables in the upstream and downstream, 
respectively, and $\mu$ is the particle's pitch angle cosine. Note that $2\mu$ in the first 
line of Equation (\ref{eqs:deltap}) is the statistical weighting factor. If there is non-zero 
background magnetic field parallel to the shock normal, a particle gains energy during its 
reflection between the upstream and downstream regions because of diffusion parallel to the 
background magnetic field. Generally, \textbf{first-order Fermi acceleration mechanism} is
not considered to have the injection problem. 
It is shown that with DSA the slope of the energy spectrum, $\gamma$, satisfies \citep[]{Decker1986}
\begin{equation}
\gamma=\frac{r+2}{2(r-1)}, \label{eqs:Gamma}
\end{equation}
where $r$ is the compression ratio across the shock.

The third mechanism is stochastic acceleration (SA) associated with the magnetic 
turbulence, also refered to as second-order Fermi acceleration \cite[]{Krulls1994, Virtanen2005}. 
The driving force of the SA process is the stochastic electric field 
$\bm{E}_b = -\bm{U}\times\delta \bm{b}$, where $\delta \bm{b}$ is the 
turbulent magnetic field superimposed on the mean magnetic field $\bm{B}$. 

With the great development of supercomputers' computational capabilities, various numerical 
simulation methods are proposed to better understand the acceleration processes of particles. 
The transport and acceleration of energetic particles in the heliosphere can be generally studied
by analyzing particle trajectories which are obtained by solving the equation of motion of particles in 
electromagnetic (EM) fields. If the EM fields are assumed to model a shock, the particle trajectory 
approach can be used to study the shock acceleration \cite[e.g.,][]{Decker1986}.
This method of numerical simulations usually requires large computational cost and 
is limited to available resources,
but if one is not interested in individual particle movements especially the details of 
particle gyromotions, the Fokker-Planck transport equation of energetic particles can be applied to 
describe the change of the distribution function \citep{Parker1965, Skilling1971, Qin2004, 
Qin2006}. \citet{Zuo2011, Zuo2013} used a computational effective method based on 
the transport equation to study the shock acceleration of energetic particles. 
However, with this method one has to assume the model of particle diffusion that is
not always established in reality.  

Charged particles can be accelerated by a shock, and in turn, shock-accelerated 
particles would influence EM fields or even excite shock waves. Therefore,
the particle acceleration and EM fields evolution have to be coupled.
Hybrid simulations are a particle-in-cell-type model that treats electrons as 
a massless fluid while treats ions kinetically \citep{Winske1996}. 
In kinetic simulations Maxwell's equations are solved self-consistently based upon plasma density and currents,
which themselves are generated from plasma particles \citep{Winske1996}.
Under certain conditions hybrid codes can be used to study particle acceleration
and shock evolution self-consistently \citep[e.g.,][]{Giacalone2004, Giacalone2005b, 
GuoFan2010, Sugiyama2011}. Hybrid codes, however, are more complicated requiring massive 
prior programming and extensive computational resources. In order to pay more attention
to the shock acceleration of energetic particles in some specific shock conditions (e.g., 
geometry, compression ratio, magnetic fields, shock speed, and magnetic
turbulence model), with more statistics but less resources needed, 
the test particle model that does not include the feedback of energetic particles 
to the EM fields, can be very useful. 
 
The diffusion of energetic particles in magnetic turbulence is very 
important to study shock acceleration of energetic particles. \citet{Jokipii1966}
developed a classical quasi-linear theory (QLT) of energetic particle’s diffusion in
a slab model of magnetic turbulence. \citet{Matthaeus2003} introduced the NonLinear
Guiding Center (NLGC) theory for the perpendicular diffusion coefficient.
\citet{Qin2007} modified the NLGC theory for perpendicular diffusion to obtain a 
NonLinear PArallel (NLPA) diffusion theory. It was shown that the solution of the 
NLGC and NLPA simultaneously agreed with simulations very well. Furthermore, 
\citet{Qin2014} obtained a NLGCE-F model by fitting the numerical results of the 
NLGC+NLPA model with polynomials. Using the model NLGCE-F one can calculate the parallel 
and perpendicular diffusion coefficients with much reduced calculations.

\textbf{Due to the effects of adiabatic deceleration, limited particle acceleration time, and
shock geometries, the energy spectrum of energetic particles in SEP events usually shows
a power law with a rollover at high energies (e.g., exponential tail) \citep{Ellison1985}
\begin{equation}
\frac{{\rm d}J}{{\rm d}E}=CE^{-\alpha}\exp\left(-\frac{E}{E_0}\right),
\label{eqs:Flux}
\end{equation}
where $C$ is a constant, $\alpha$ is the spectral index, and $E_0$ is the bend-over energy.} 

In the present paper, we study the parallel shock acceleration of test particles 
by numerically solving the equation of motion of particles in the shock frame in 
prescribed turbulent magnetic field. By calculating the trajectories of a large number 
of particles we obtain power-law energy spectra of accelerated particles with a 
bend-over energy $E_0$. We then investigate the variation of $E_0$ over time and 
compare the simulations with the results from theoretical models.
We describe our simulation model in section 2, followed by the theoretical models of 
the bend-over energy in section 3. In section 4, we present the numerical results and 
the comparison with different models. Finally, we show conclusions and discussions 
in section 5.

\section{Simulation Models}
The test-particle trajectory in the shock frame is controlled by the equation of motion  
\begin{equation}\label{eqs:Newton}
\frac{{\rm{d}}\bm{p}}{{\rm{d}}t}=q\left(\bm{E}+\bm{v}\times\bm{B}\right),
\end{equation}
where $\bm{p}$, $\bm{v}$, and $q$ are the particle momentum, velocity, and
electric charge, respectively, and $\bm{E}$ and $\bm{B}$ are the local electric 
field and magnetic field, respectively. 
We simplify the shock as an infinite plane 
\citep{KongEA2017,QinEA2018}. The shock normal is in the direction anti-parallel to 
the $z$ axis, and the shock lies in the $x-y$ plane.
The upstream and downstream plasma speed, $\bm{U}_1$ 
and $\bm{U}_2$, are both assumed to be parallel to the shock normal. The average 
magnetic fields in the upstream and downstream regions, are $\bm{B}_{01}$ and 
$\bm{B}_{02}$, respectively. 

We use a composite slab plus two-dimensional (2-D) model of static turbulent magnetic 
field \citep{MatthaeusEA90,Mace2000, Qin2002a, Qin2002b}, which is different from that in
\citet{Decker1986}.
The total magnetic field is written as
\begin{equation}\label{eqs:Btotal}
\bm{B}(x^\prime, y^\prime, z^\prime)=\bm{B}_0(z^\prime)+
\bm{b}(x^\prime, y^\prime, z^\prime), 
	\nonumber
\end{equation}
where a Cartesian coordinate system is adopted, and the $z^\prime$-axis is 
in the direction parallel to the mean magnetic field $\bm{B}_0$. The turbulent 
magnetic fluctuations, $\bm{b}$, are composed of a slab and 2D component,
\begin{equation}
\bm{b}(x^\prime, y^\prime, z^\prime)=\bm{b}_{slab}(z^\prime)+
\bm{b}_{2D}(x^\prime,y^\prime),\label{eqs:Btubulence}
\end{equation}
where both of the two components are perpendicular to $\bm{B}_0$.

The induced electric field $|\bm{e}_k|\approx V_A|\bm{b}_k|$ of the turbulent magnetic 
field $\bm{b}_k(z)$ in the plasma frame is neglected because of the fact that the 
Alfv$\rm\acute{e}$n speed is far lower than the speed of particles ($V_A \ll |\bm{v}|$) \citep{Jokipii1971,Decker1986}. %
Therefore, there is no electric field in the plasma frame on each side of the shock plane,
and in the shock frame of reference the electric field $\bm E$ is
from the convection due to the plasma bulk speed.
In our simulations, the turbulence has a Kolmogorov's spectrum with
a power 
index of $\nu=-5/3$ at the high wavenumber $k$ for each component. 
Except the shock compression ratio $r=3.85$, the other background
parameters are chosen as those in \citet[]{Decker1986}. 
In the upstream, the mean magnetic field is $B_{01}= 50$ G, the Alfv$\rm\acute{e}$n speed
is $V_{A1}=1.1\times10^6~\rm{m/s}$,
and the bulk velocity is $U_1=3.3\times 10^6~\rm{m/s}$. The correlation length of 
the slab component of the turbulent magnetic field is 
$\lambda=6.67\times 10^{-9}~\rm{au}=1.00\times 10^{3}~{\rm m}$. 
The turbulent magnetic field $\bm{b}_{slab}(z^\prime)$ is created with 
fast Fourier transform (FFT) in a box with a size $L_{z^\prime}=64\lambda$ 
along the mean magnetic field $\bm{B}_0$ and 
number of grids $N_{z^\prime}=2^{22}=4194304$. In addition, to be different from that 
in \citet[]{Decker1986}, we apply the 2D component 
magnetic turbulence with the 2D
correlation scale $\lambda_x=\lambda/2.6$ \citep{OsmanAHorbury07, 
DoschEA13, WeygandEA09, WeygandEA11, ShenAQin18}, in a box with a size
$L_{x^\prime}=L_{y^\prime}=10\lambda$ 
and number of grids $N_{x^\prime}=N_{y^\prime}=4096$. The energy ratio of the two components 
of turbulence is taken to be $E_{slab}:E_{2D}=1:4$. The turbulent level, $(b/B_0)^2$,
is set 0.19 and 0.38 in the up- and down-stream, respectively.
In the downstream side, the bulk velocity $\bm{U}_2$ and mean magnetic field $\bm{B}_{02}$
are 
obtained from the Rankine-Hugoniot relations. Since the average of turbulence magnetic field 
is zero and turbulence level is small, we do not consider the Rankine-Hugoniot 
relations for the turbulent magnetic field in the downstream of the shock.

We adopt a numerical code developed by \citet{Sun2007} (also used in \citet{KongEA2017}
and \citet{QinEA2018}) using an adjustable time step 
fourth-order Runge-Kutta method with an accuracy of $10^{-9}$ to obtain test particle
trajectories by solving the equation of motion of particles, Equation (\ref{eqs:Newton}).
A total of 60,000 protons 
with an initial energy of $E_{in}=30$ keV in the shock frame are isotropically injected
in the upstream with a distance $d=1.1 r_g$ to the shock front, where $r_g$ is the 
proton gyroradius. \textbf{The simulations of test particles are performed with an 
longest acceleration time of $t_{acc}=500~\rm{ms}$ without space limitations.}

\section{Theoretical Models}

\subsection{Models of Diffusion}
\begin{figure}
\epsscale{1.}
\plotone{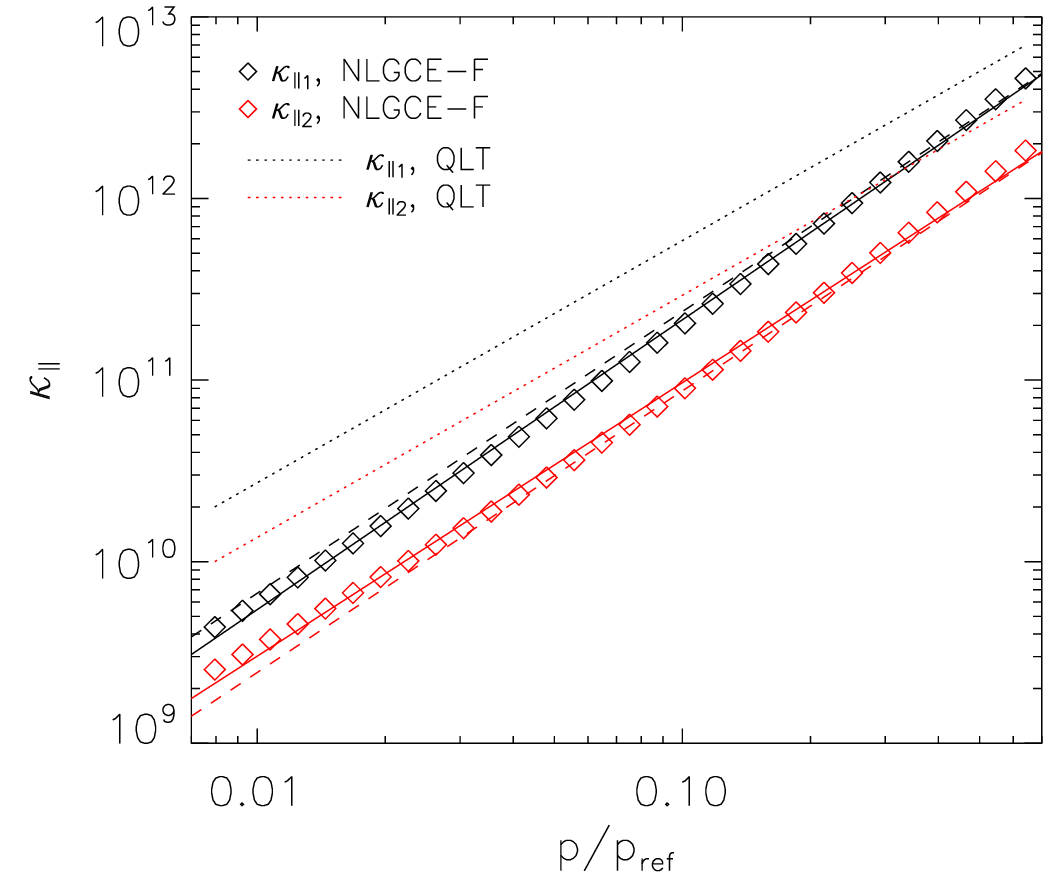}
        \figcaption{Parallel diffusion coefficients, $\kappa_{\parallel 1}$ 
        and $\kappa_{\parallel 2}$, 
in the upstream (black) and downstream (red) of the shock, respectively, 
as a function of particle momentum $p/p_{ref}$, where $p_{ref}$ 
is the momentum of a proton with rigidity $R=1$ GV.
The diamonds represent calculation results from NLGCE-F. The solid
lines indicate the fitting of the NLGCE-F results using the power-law form in
Equation (\ref{eqs:kappai}) with power indice 
$\xi_1$ and $\xi_2$ in the upstream and downstream, respectively. To replace
the power indice $\xi_1$ and $\xi_2$ with the average value, the solid lines
are changed to the dashed lines. The dotted lines indicate the QLT results.
\label{fig:kappafit}}
\end{figure}

 On the basis of the 
nonlinear diffusion theory \citep[NLGCE-F]{Qin2014}, 
{\bf we can calculate $\kappa_\parallel$ using the computer code downloaded 
from the website \url{http://www.qingang.org.cn/code/NLGCE-F} for different 
proton momentum in both up- and down-stream.
Besides, using quasi-linear theory (QLT) \citep{Jokipii1966},
we have \citep[see also, e.g.,][]{Qin2002PhD}
\begin{equation}
\kappa_\parallel=1.72\frac{\lambda^{2/3}B_0^{-1/3}}
{(b_{slab}/B_0)^2}\left(\frac{R}{c}\right)^{1/3}v, \label{eqs:kappapara}
\end{equation}
considering the spectral index of slab turbulence in the inertial range being $5/3$.
According to QLT with Equation (\ref{eqs:kappapara}), it can be assumed that
diffusion coefficients $\kappa$ can be written as
\begin{equation}\label{eqs:kappai}
\kappa=\kappa_R\left(\frac{p}{p_{ref}}\right)^\xi, \label{eqs:kappaRxi}
\end{equation}
where $\kappa_{R}$ and $\xi$ are constants, and 
$p_{ref}=5.34\times10^{-19}~{\rm kg\cdot m/s}$ is the momentum of a
proton with rigidity $R=1$ GV.

Figure \ref{fig:kappafit} shows parallel diffusion coefficients, 
$\kappa_\parallel$ as a function of particle momentum $p/p_{ref}$, where
$p_{ref}$ is the momentum of a proton with rigidity $R=1$ GV.
Here we show variables in up- and down-stream with subscripts $i=1$ and $2$,
color black and red, respectively.
The diamonds represent calculation results from NLGCE-F. The solid
lines indicate the fitting of the NLGCE-F results using the power-law form in
Equation (\ref{eqs:kappai}) with parameters $\xi_i$ and $\kappa_{Ri}$. 
To replace the power indice $\xi_1$ and $\xi_2$ with the average value, 
$(\xi_1+\xi_2)/2$, the solid lines are changed to the dashed ones. 
It is shown that the dashed lines agree approximately with the solid ones, so
one can adopt the average value $\xi$ for both up- and down-stream in
NLGCE-F. In addition, the dotted lines indicate
the QLT results with parameters $\xi_i$ and $\kappa_{Ri}$ in Equation 
(\ref{eqs:kappai}) obtained from the formula (\ref{eqs:kappapara}) 
analytically. Here for QLT, $\xi=\xi_1=\xi_2$. The values 
of $\xi_i$ and $\kappa_{Ri}$ from NLGCE-F and QLT are listed in Table 
\ref{theoryPara}.}

\subsection{Model of The Bend-Over Energy}

It is very interesting to study the time evolution of the bend-over energy 
if we assume SEPs are accelerated by a shock.
Using the DSA model \citep{Drury1983} for each shock crossing of particles,
from Equation (\ref{eqs:deltat}) and Equation (\ref{eqs:deltap}), 
one can obtain
\begin{equation}
\frac{{\rm{d}}p}{{\rm{d}}t}\approx\frac{\Delta p}{\Delta t}=
	\frac{1}{3}\left(U_1-U_2\right)\left(\frac{\kappa_1}{U_1}+
	\frac{\kappa_2}{U_2}\right)^{-1}p.\label{eqs:dpdt}
\end{equation} 
As seen from Equation (\ref{eqs:dpdt}), the acceleration rate of particles by
the shock depends on the diffusion coefficient $\kappa_i$, which allows us to  
get different models of shock acceleration rate with different diffusion models.

The momentum, $p_{acc}$, of accelerated particles is obtained 
by integrating Equation (\ref{eqs:dpdt}) considering Equation
(\ref{eqs:kappaRxi}),
\begin{equation}
\left(\frac{p_{acc}}{p_{ref}}\right)^{\xi_1}+g\left(\frac{p_{acc}}
{p_{ref}}\right)^{\xi_2}=\left(\frac{p_0}{p_{ref}}\right)^{
\xi_1}+g\left(\frac{p_0}{p_{ref}}\right)^{\xi_2}+
\frac{1}{3}\frac{
U_1^2}{r\kappa_{R1}}\xi_1(r-1)t,\label{eqs:ptimplicant}
\end{equation}
where $g=\xi_1\kappa_{R2}r/\left(\xi_2\kappa_{R1}\right)$, $p_0$ is 
the particle initial momentum, and the corresponding energy $E_{acc}$ is
\begin{equation}
E_{acc}=\sqrt{p_{acc}^2c^2+E_p^2}-E_p, \label{eqs:Et}
\end{equation}
where $E_p$ is the static energy of a proton. 
The Equation (\ref{eqs:ptimplicant}) is an implicant, by numerically solving
which the particle momentum $p_{acc}$ with time $t$ of shock acceleration
could be obtained. It is less possible for particles to be accelerated to
energies higher than $E_{acc}$. Therefore, the energy spectrum of particles
accelerated by a shock would turn over at the energy above $E_{\rm acc}$,
and we can define a bend-over energy, $E_0$, as a function of time $t$
\begin{equation}
E_0\equiv E_{\rm acc}=\sqrt{p_{acc}^2c^2+E_p^2}-E_p.
\label{eqs:Eb}
\end{equation}
If $\xi_1=\xi_2\equiv\xi$, the Equation (\ref{eqs:ptimplicant}) could be 
solved directly as,
\begin{equation}
p_{acc}=p_{ref}\left[\left(\frac{p_0}{p_{ref}}\right)^{\xi}+
\frac{U_1^2\xi}{3}\frac{r-1}{r(\kappa_{R1}+r\kappa_{R2})}t\right]^{1/\xi}.
\label{eqs:pt}
\end{equation}

{\bf Since we study parallel shock acceleration in this work, diffusion 
coefficients $\kappa_1$ and $\kappa_2$ in Equation (\ref{eqs:dpdt}) 
refer to parallel diffusion.
Note that for QLT, we can directly get 
$p_{acc}$ from Equation (\ref{eqs:pt}) due to the fact that the power
indice in the up- and down-stream are the same. Whereas for NLGCE-F, $\xi_1$ in 
the upstream is not equal to $\xi_2$ in the downstream. 
In order to get an explicit expression of particle momentum $p_{acc}$ in Equation
(\ref{eqs:ptimplicant}), we assume $\xi$ as $(\xi_1+\xi_2)/2$.}
Therefore, the momentum of accelerated particles $p_{acc}$ with QLT and
NLGCE-F can be calculated directly from Equation (\ref{eqs:pt}), and the 
corresponding bend-over energy, which are indicated by $E_0^{\text{QLT}}$ and
$E_0^{\text{NLGCE-F}}$, respectively, could be obtained.

\subsection{Energy Spectrum Power Law Index}
As discussed in \citet[]{Decker1986}, if charged particles are injected near
the shock and accelerated for a time long enough, the energy spectrum
becomes stable with
a power law index $\gamma$ satisfying Equation (\ref{eqs:Gamma}).
In this work, the compression ratio of the shock is $r=3.85$, so the energy spectrum 
of the accelerated particles, has a theoretical 
power law index of $\gamma=1.03$. The spectral index from simulations, $\alpha$, in 
Equation (\ref{eqs:Flux}) in the energy range below the bend-over energy, 
can be compared with the theoretical power law index $\gamma$.
 
\section{Numerical Results and Comparison with Theories}

\begin{figure}
\epsscale{1.}
\plotone{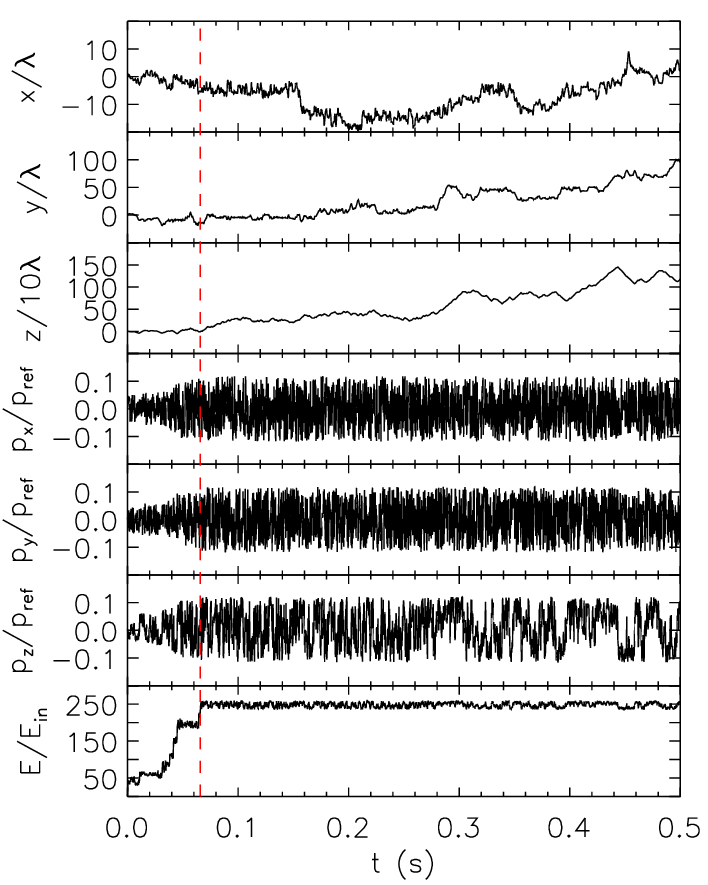}
        \figcaption{Trajectory of a test particle as a function of time. The top three panels show
the particle position in the Cartesian coordinate system, the fourth to sixth panels show the particle
momentum in each direction, and the bottom panel indicates the particle energy. The vertical red line
indicates the time when the particle does not cross the shock again.\label{fig:track}}
\end{figure}

Figure \ref{fig:track} illustrates the trajectory of one of the test particles 
accelerated by a parallel shock as a function of time. The top three panels show
the $x-$, $y-$, and $z-$ components, respectively, of the particle position 
in units of $\lambda$. The fourth to sixth panels show the $x-$, 
$y-$, and $z-$ components, respectively, of the particle momentum in units of $p_{ref}$. 
The bottom panel shows the particle energy in units of its initial energy $E_0=30$ keV. 
Generally, a particle will get accelerated and gain energy 
when it crosses the shock plane back and forth. From the top three panels of 
Figure \ref{fig:track} we can see that within the initial $\sim$ 0.06 s (vertical
dashed line), the particle crosses the shock plane many times, but beyond $\sim$ 
0.06 s the particle does not cross the shock any more, as the $z-$ component of 
particle position increases persistently with time, indicating the particle moves 
far away from the shock plane. In addition, from the 4th to 6th panels of Figure 
\ref{fig:track}, one can find that the particle generally performs gyro-rotation 
in the $x-y$ plane. 
It can be seen that within $\sim$ 0.06 s the crests of $p_x$ and $p_y$ increase 
much more significantly than the crests of $p_z$, suggesting that the energy 
gain is mainly in the gyro-rotation plane during the shock crossings. However, 
beyond $\sim$ 0.06 s, the crests of $p_x$ and $p_z$ do not increase anymore, 
but the magnitude of $p_z$ increases to the similar level of $p_x$ and $p_z$.
It is assumed that the energy homogenization among different directions are 
due to the pitch angle diffusion by magnetic turbulence. From the bottom panel 
of Figure \ref{fig:track}, the particle gains energy more than 200 times of 
its initial energy. But beyond $\sim$ 0.06 s, the particle energy keeps almost constant.

\begin{figure}
\epsscale{1.}
\plotone{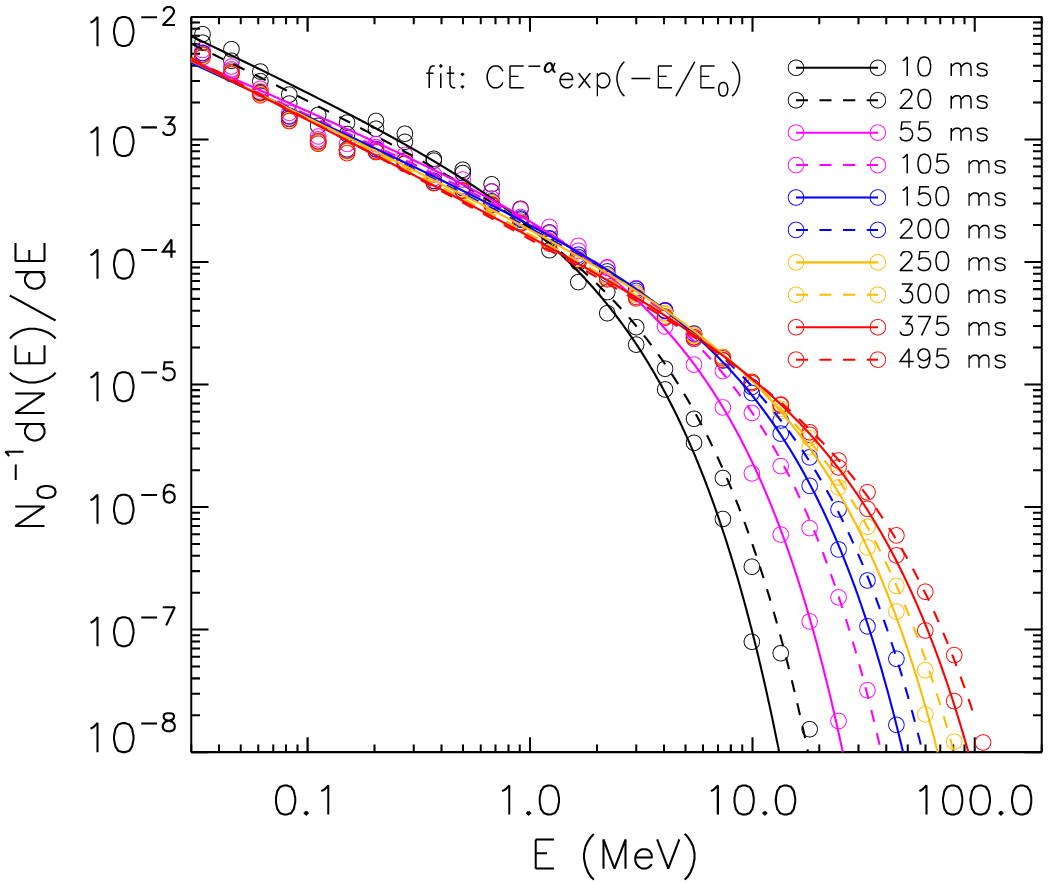}
        \figcaption{Downstream energy spectra of accelerated
particles for different simulation times (circles). Solid and dashed lines indicate Fits to the 
simulated energy spectra using the function form in Equation \ref{eqs:Flux},
at various simulation times.
\label{fig:fluxt1}}
\end{figure}

From the trajectories of test particle simulations, we calculate the energy spectra of
accelerated particles for different simulation time.
\textbf{In Figure \ref{fig:fluxt1} circles show the energy spectra of accelerated 
particles with different simulation time in the downstream of the shock. Note that 
we do not show the 
energy spectra with energies lower than the initial energy $E_{in}$, since in reality 
below $E_{in}$ the background spectrum is dominant. Here we show spectra of the 
acceleration time of $10$, $20$, $55$, ..., $495$ ms. We can see that the energy
spectrum hardens and extends 
to higher energies with increasing of time. The spectrum at $t=495$ ms reaches as
high as $100$ MeV
compared to $\sim 10$ MeV for the case of $t=10$ ms. In addition, the spectra at
lower energies show a power-law with a bend-over energy $E_0$. As time increased the 
spectra with power-law do not change significantly with the bend-over energy increased.}

\textbf{Furthermore, we fit the simulated data 
using a power law with an exponential tail from Equation (\ref{eqs:Flux}) in 
log-log space, adopting the nonlinear least-squares fitting algorithm.
The best-fit parameters, $C$, $\alpha$, and $E_0$ are listed in Table \ref{powerlawExp}.
In Figure \ref{fig:fluxt1}, we plot the fitted energy spectra with solid and dashed
lines. 
The result provides a good fit to the simulated energy spectra. This indicates
that in our simulations the spectra of the accelerated particles in the downstream exhibit
a form of a power law with an exponential tail.} 

\begin{figure}
\epsscale{1.0}
\plotone{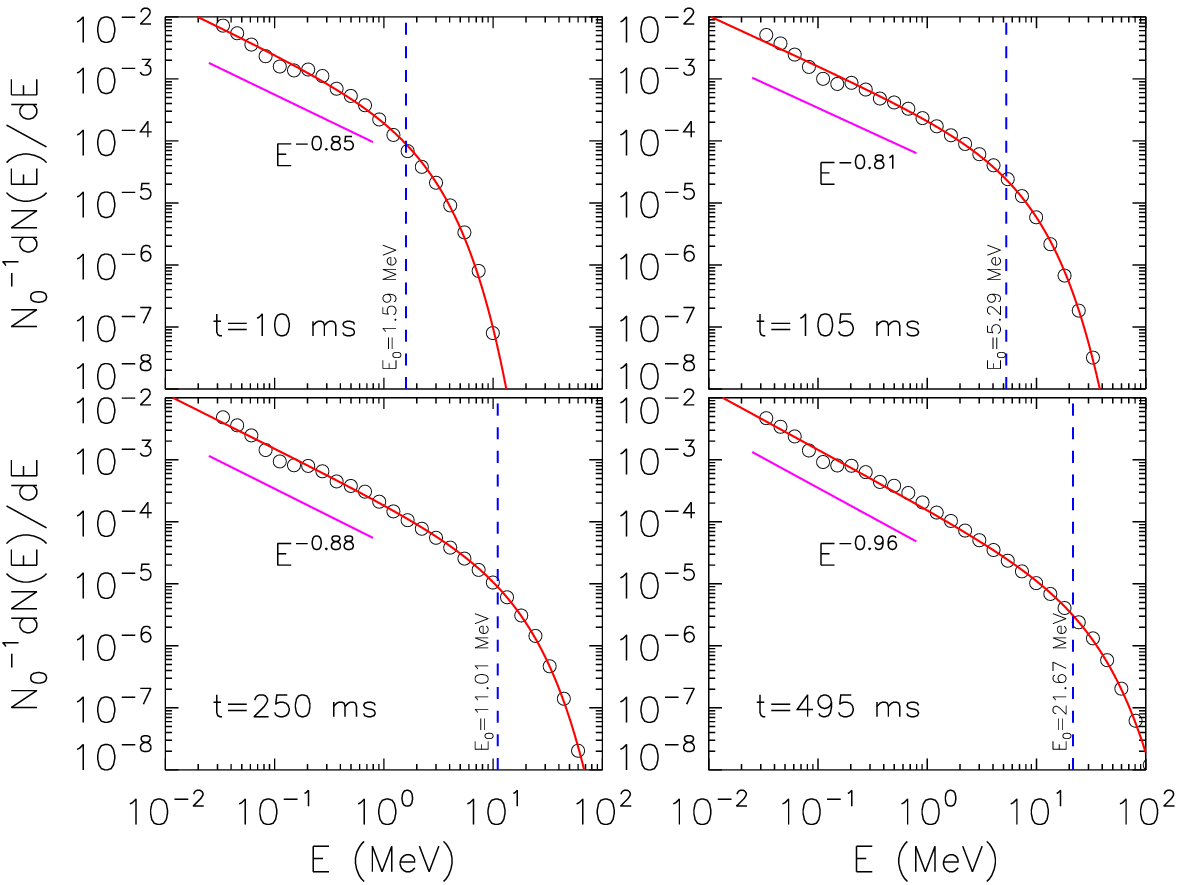}
        \figcaption{Fits to the simulated spectra (open circles) 
at $t=10$, $105$, $250$, $495$ ms are plotted in red curves. The
blue vertical line and magenta oblique line in each panel denote the bend-over 
energy $E_0$ and the spectral index.
\label{fig:fluxt2}}
\end{figure}

\textbf{ In addition, we present the energy spectra 
of $t=25$, $90$, $225$, $440$ ms in Figure \ref{fig:fluxt2}. It is shown that the power law
with an exponential tail form fits well to the simulated energy spectra. The oblique and
vertical lines indicate the spectral index and bend-over energy. From
this figure we can see that the bend-over energy, $E_0$, increases from $\sim1.6$ MeV to 
$\sim 22$ MeV within $500$ ms. Moreover, in the same acceleration time range, generally,
spectral index $\alpha$ increases from $0.85$ to $0.96$.}

\begin{figure}
\epsscale{1.}
\plotone{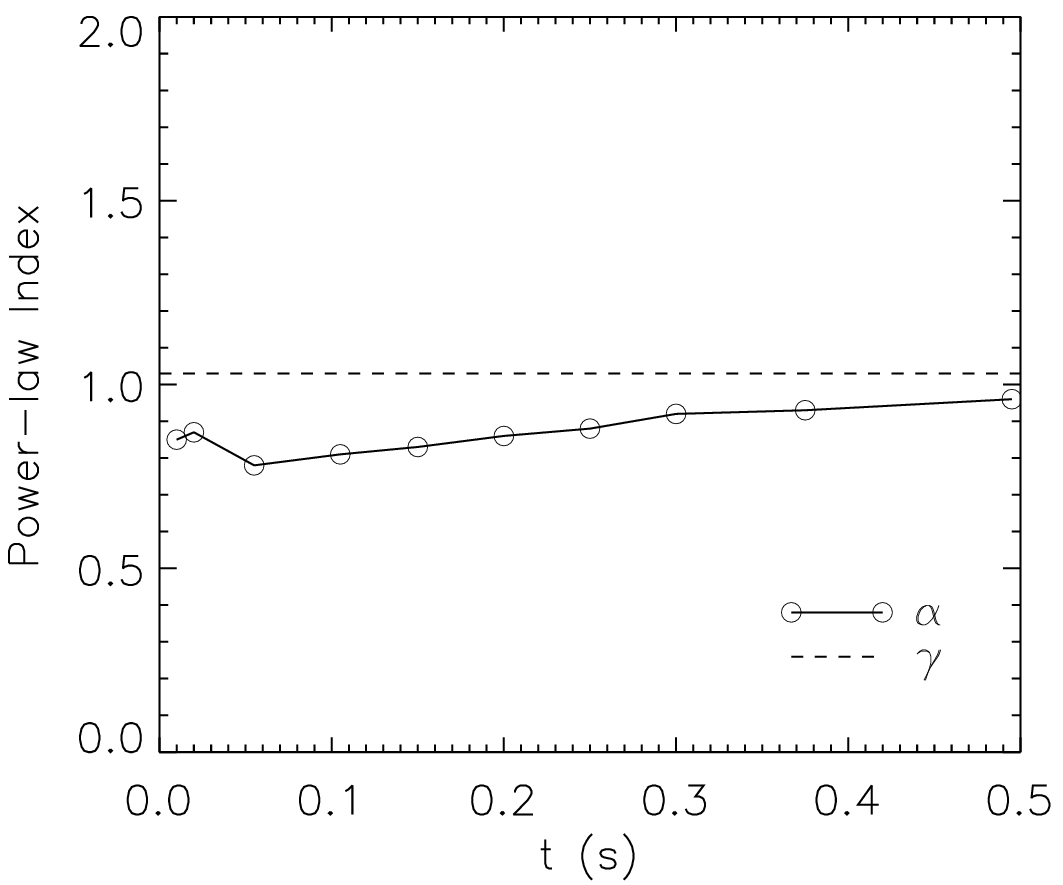}
        \figcaption{Spectral index of shock accelerated particles as a function of time. The circles 
indicate results from simulations, and the dashed line corresponds to DSA theory.  \label{fig:slope}}
\end{figure}

\textbf{In Figure \ref{fig:slope} we show the evolution of the power-law index, $\alpha$,
of the shock accelerated particle energy spectrum from simulations as shown in Figure
\ref{fig:fluxt1} with time (open circles). The dashed line indicates 
the theoretical index $\gamma$ from Equation (\ref{eqs:Gamma}) by \citet[]{Decker1986}.
It is shown that the power law index from simulations is always smaller than the  
theoretical results, but generally it increases with acceleration time. In addition, as
the acceleration time becomes larger, the simulated power-index approaches the theoretical
one more and more. }

\begin{figure}
\epsscale{1.}
\plotone{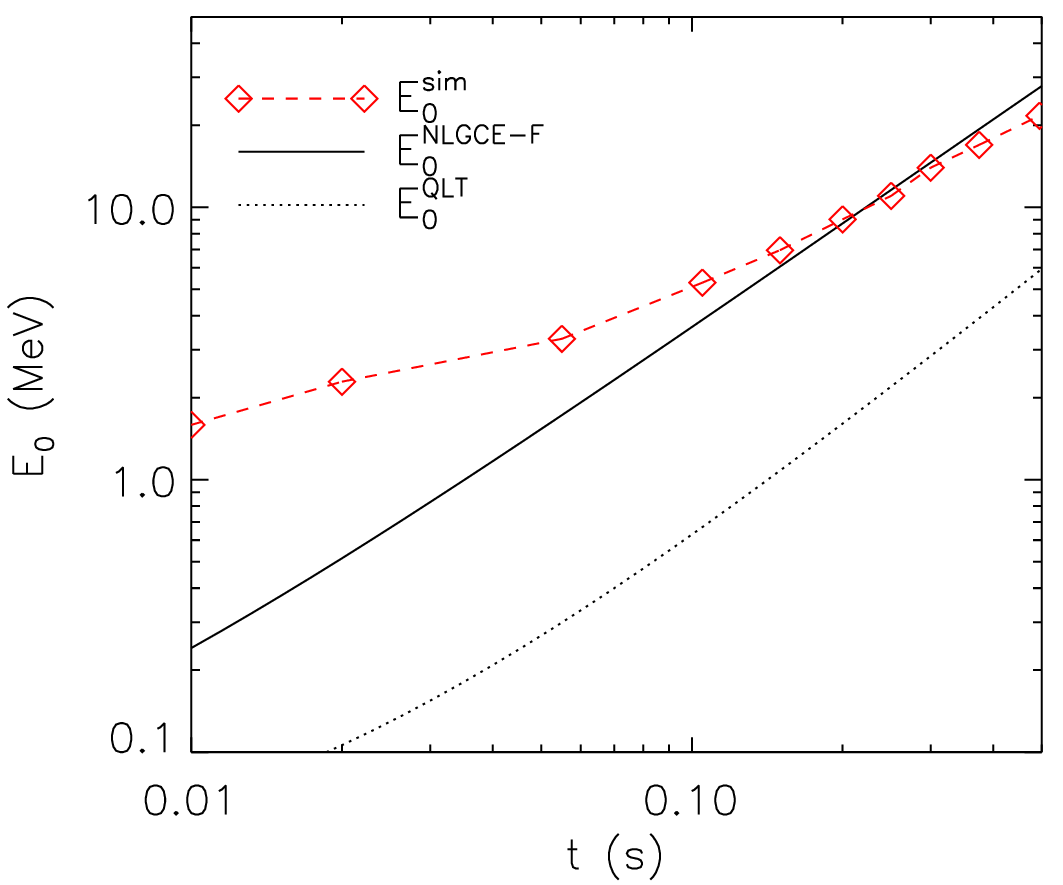}
        \figcaption{Bend-over energy, $E_0$, as a function of time from 
simulations (diamonds), theory with NLGCE-F (solid line), and theory with QLT
(dotted line).  \label{fig:E0t}}
\end{figure}

\textbf{Figure \ref{fig:E0t} shows the time evolution of the bend-over energy, $E_0$. Red
diamonds show the results of bend-over energy from simulations $E_0^{\text{sim}}$. Solid 
and dotted lines indicate the results from theory Equation (\ref{eqs:pt}) with diffusion
models NLGCE-F
and QLT, $E_0^{\text{NLGCE-F}}$ and $E_0^{\text{QLT}}$, respectively.
It can be seen that $E_0^{\text{sim}}$, $E_0^{\text{NLGCE-F}}$, and $E_0^{\text{QLT}}$
all increase with time.
In addition, $E_0^{NLGCE-F}$ is always about $5$ times that of $E_0^{QLT}$.
At short acceleration time, $t\sim 10$ ms, $E_0^{sim}$ is about $6.5$ times that of 
$E_0^{NLGCE-F}$, but $E_0^{sim}$ increases slower than $E_0^{NLGCE-F}$ does, so at 
$t\gtrsim 100$ ms $E_0^{sim}$ is similar to $E_0^{NLGCE-F}$. In conclusion, at any
acceleration time, $E_0^{NLGCE-F}$ is more consistent with $E_0^{sim}$ than $E_0^{QLT}$
does. Additionally, at larger acceleration time, $E_0^{NLGCE-F}$ agrees well with 
$E_0^{sim}$.}

\section{Summary and Conclusions}  

The well-known hybrid code is an effective approach to 
study particle acceleration and shock evolution self-consistently.
It is, however, of interest to study the shock acceleration of energetic particles 
with some pre-determined shock conditions, e.g., geometry, compression ratio, 
magnetic fields, shock speed, and magnetic turbulence models.
In addition, since test-particle simulations require much less computational resources, 
it is possible for us to study particle acceleration with more statistics in larger computation space and time. 
In this paper, we focus on particle acceleration at the parallel shock
with numerical calculations of the trajectories of test particles
by solving the equation of motion in turbulent magnetic fields.
We simplify the shock to be an infinitely thin plane, and assume static magnetic 
turbulence superimposed on the static background magnetic field.
 
Our simulations indicate that charged particles can be accelerated to high energies, 
and even the energy gains of some particles can reach a few hundreds of times of their 
initial energy within several hundred milliseconds. We find that the energy spectrum of 
the shock-accelerated particles from simulations shows \textbf{a power law with an 
exponential tail.}  The spectral index of the lower energy range agrees well with the 
theoretical power law index model by \citet[]{Decker1986}.
It is also found that from simulations the bend-over energy increases with time.

Theoretically, using the mean acceleration time and the average momentum change during
each cycle of 
particles crossing of the shock \citep{Drury1983}, we get the energy $E_{acc}$ of 
shock-accelerated particles as a function of time. In addition, 
we assume the bend-over energy is equal to $E_{acc}$ since it is less possible 
for particles to be accelerated to energies higher than $E_{acc}$ at any time. 
The theoretical bend-over energy depends on the diffusion model. We compare the 
bend-over energy from simulations with that from theory adopting either QLT or 
NLGCE-F for the diffusion model, and find that \textbf{the simulated bend-over energy
has a good agreement 
with the bend-over energy model obtained with NLGCE-F.} 
This result implies that the mothod of our test-particle simulations for shock
accelerations can indirectly demonstrate that the diffusion model NLGCE-F is
more accurate than QLT \citep{Qin2014}.

Our results from both numerical simulations and theoretical models show that the
\textbf{charged particles accelerated at the parallel shock could produce the energy spectrum
of a power-law with an exponential tail.} 
In this work, because of the limit of computational resources, we do not simulate
the shock acceleration for a very long time, so that we do not obtain a power-law
spectrum of energetic particles with the bend-over energy as large as that observed
in the solar wind. In the future, we plan to study shock acceleration with much longer
time so that we might be able to obtain more realistic power-law spectrum of energetic particles. 

\acknowledgments
This work was supported by the Strategic Priority Research Program
of Chinese Academy of Sciences (Grant No. XDA17010301). The work was
also supported, in part, under grants NNSFC
41874206 and NNSFC 41574172.
The work was carried out at National Supercomputer Center 
in Tianjin, and the calculations were performed on TianHe-1 (A).

\clearpage
\begin{table}[ht]
\begin{center}
\caption{Values of $\xi_i$ and $\kappa_{Ri}$ from NLGCE-F and QLT
	\label{theoryPara}}
\begin{tabular}{cccc}
\hline\hline
Theory & Parameter & Upstream & Downstream \\
\tableline
\multirow{2}{*}{NLGCE-F} & $\xi_i$ & $1.60$ & $1.51$ \\
\cline{2-4}
 & $\kappa_{Ri}~({\rm m^2/s})$ & $8.48\times10^{12}$ & $3.10\times10^{12}$ \\
\tableline
\multirow{2}{*}{QLT} & $\xi_i$ & $1.33$ & $1.33$ \\
\cline{2-4}
 & $\kappa_{Ri}~({\rm m^2/s})$ & $1.26\times10^{13}$ & $6.32\times10^{12}$ \\
\tableline
\end{tabular}
\end{center}
\end{table}
\clearpage

\begin{table}[ht]
\begin{center}
\caption{Fitted Parameters for Time-dependent Energy Spectra
 	\label{powerlawExp}}
\begin{tabular}{cccc}
\hline\hline
t (ms) & $C(10^{-3})$ & $\alpha$ & $E_0$ (MeV)  \\
\tableline
10 & 0.36 & 0.85 & 1.59 \\
20 & 0.29 & 0.87 & 2.29 \\
55 & 0.29 & 0.78 & 3.29 \\
105& 0.25 & 0.81 & 5.29 \\
150& 0.23 & 0.83 & 6.96 \\
200& 0.21 & 0.86 & 9.03 \\
250& 0.20 & 0.88 & 11.01\\
300& 0.18 & 0.92 & 13.98\\
375& 0.17 & 0.93 & 16.95\\
495& 0.16 & 0.96 & 21.67\\
\tableline
\end{tabular}
\end{center}
\end{table}
\clearpage

\clearpage

\clearpage

\clearpage

\clearpage

\end{document}